\documentclass{article}   

\def\mypagenumber{1}
\def\mydate{September 19, 1997}

\date{}
\def\mydate{19 September 1997}

\newcommand{\beeq}{\begin{equation}}
\newcommand{\eneq}{\end{equation}}
\newcommand{\beqn}{\begin{eqnarray}}
\newcommand{\eeqn}{\end{eqnarray}}

\def\mybig{\displaystyle \strut }

\def\dd{\partial}
\def\la{\raise.16ex\hbox{$\langle$} \, }
\def\ra{\, \raise.16ex\hbox{$\rangle$} }
\def\go{\rightarrow}

\def\onehalf{ \hbox{${1\over 2}$} }

\def\psibar{ \psi \kern-.65em\raise.6em\hbox{$-$} }
\def\lowpsibar{ \psi \kern-.65em\raise.6em\hbox{$-$}\lower.6em\hbox{} }
\def\tpsi{ \widetilde\psi}

\def\tpsibar{ \tpsi \kern-.65em\raise.9em\hbox{$-$}\lower.9em\hbox{} }
\def\zetabar{ \zeta\kern-.65em\raise.6em\hbox{$-$}\lower.6em\hbox{} }
\def\mbar{ m \kern-.78em\raise.4em\hbox{$-$}\lower.4em\hbox{} }

\def\Bbar{ B \kern-.73em\raise.6em\hbox{$-$}\hbox{} }

\def\L{ {\cal L} }

\def\boxit#1{$\vcenter{\hrule\hbox{\vrule\kern3pt
     \vbox{\kern3pt\hbox{#1}\kern3pt}\kern3pt\vrule}\hrule}$}
\def\bigbox#1{$\vcenter{\hrule\hbox{\vrule\kern5pt
     \vbox{\kern5pt\hbox{#1}\kern5pt}\kern5pt\vrule}\hrule}$}

\def\@makefnmark{{$\!^{\@thefnmark}$}}

\begin{document}

\setcounter{page}{\mypagenumber}

{\baselineskip=10pt \parindent=0pt \small
{\mydate \hfill UMN-TH-1608/97\\} 
}

\vskip .5cm

\leftline{\huge Gauge Theory Description of Spin}
\vskip .1cm
\leftline{\huge Chains and Ladders}

\vskip .7cm

\leftline{\Large Yutaka Hosotani}

\vskip .7cm

\begin{center}
{\begin{minipage}{10.2truecm}
                 \parindent=0pt  \small
ABSTRACT ~
An $S={1\over 2}$ anti-ferromagnetic spin chain is mapped to the two-flavor
massless Schwinger model, which admits a gapless mode.
In a spin ladder system rung interactions break the chiral invariance.
These systems are solved by bosonization.  If the number of legs 
in a cyclically symmetric ladder system is even, all of the gapless modes of
spin chains become gapful.  However, if the number of legs is
odd,  one combination of the gapless modes remains gapless.  For a 
two-leg system we find that the spin gap is about  $ .36 \, |J'|$
when the inter-chain Heisenberg coupling $J'$ is  small compared with
the intra-chain Heisenberg coupling.
\par
                 \end{minipage}}\end{center}

\vskip .6cm

An anti-ferromagnetic spin chain is defined by
\beeq
H_{\rm chain}[\vec S] = J \sum \vec S_n \cdot \vec S_{n+1}  
\qquad (J>0)~.
\label{chain1}
\eneq
$n_\ell$ spin chains coupled  by a rung interaction define an 
$n_\ell$-leg spin ladder:
\beeq
H_{\rm ladder} = \sum_{j=1}^{n_\ell} H_{\rm chain}[\vec S^{(j)}]
     +  J' \sum_{(jk)}   \vec S_n^{(j)} \cdot \vec S_n^{(k)} ~.
\label{ladder1}
\eneq
$j$ and $k$ are leg indices.  In experimental samples the rung
interaction extends over $(jk)=(12), (23), \cdots, (n_\ell$-$1, n_\ell)$.
In a cyclically symmetric ladder which we shall discuss below, an
interaction over $(jk)=(n_\ell,1)$ is also added.

For $S=\onehalf$ the spin chain Hamiltonian (\ref{chain1}) is exactly 
solved by the Bethe ansatz.\cite{Bethe}  It has a gapless excitation.  In
experiments it is observed that a two-leg ladder system has no gapless
excitation,  whereas a three-leg ladder system remains gapless.\cite{exp}
The theme of this report is to show how the gapless modes
of spin chains become gapful when the number $n_\ell$  of legs in the
ladder system is even.\cite{Hosotanispin}

Haldane conjectured that the spectrum of the spin chain system is gapless
or gapful for a half-odd-integer or integer spin $S$, respectively.
The conjecture has been supported both experimentally and
theoretically.  In a ladder system with $S=\onehalf$ it is believed
that the spectrum is gapless or gapful for an odd or even $n_\ell$, 
respectively.  However, a subtle issue remains as the spectrum may depend
on the topological structure and magnitude of the rung interaction in
(\ref{ladder1}).


To analyse (\ref{ladder1}) we first establish the equivalence
between the $S=\onehalf$ spin chain (\ref{chain1}) and the two-flavor
Schwinger model.  In the literature spin systems are often mapped to
non-linear sigma models which is justified for large $S$.\cite{sigma}  There
is an alternative mapping which works well for $S=\onehalf$.

Write  $\vec S_n = c_n^\dagger \onehalf \vec \sigma c_n^{}$. 
After reshuffling electron operators in $\vec S_n \vec S_{n+1}$ and
performing the  Hubbard-Stratonovich transformation, 
the system (\ref{ladder1}) is converted to a Lagrangian
\beeq
\sum \Big\{ ~ i \hbar c_n^\dagger \dot c_n^{} 
   - \lambda_n (c_n^\dagger c_n^{} - 1)    
-{J\over 2} (U_n^* U_n - U_n^{} c_n^\dagger c_{n+1}^{} - 
U_n^* c_{n+1}^\dagger c_n^{} ) 
~ \Big\} ~.
\label{chain2}
\eneq
The Lagrange multiplier $\lambda_n$  enforces the half-filling condition at
each site.  Fluctuations of link variables $U_n$'s  are large in phase,
but their magnitude is frozen about $1/\pi$.  One can write
$U_n =  e^{i \ell A_n}/\pi$ where $\ell$ is the lattice spacing.
(\ref{chain2}) thus becomes lattice QED.

We take the continuum limit.  Note that in an antiferromagnetic chain
two sites form a fundamental block. The even-odd site index becomes a spin
index of the Dirac field, whereas the original electron
spin index becomes a flavor index, $a$, as $U_n$ couples to the spin singlet
combination $c_n^\dagger c_{n+1}^{}$  in (\ref{chain2}).   
In the continuum limit
\beeq
\L_{\rm chain} = - {1\over 4e^2} \, F_{\mu\nu}^2
+ \sum_{a=1}^2 c \lowpsibar^{(a)}  \gamma^\mu 
  \Big( i \hbar \dd_\mu - {1\over c} A_\mu \Big)
\psi^{(a)} + {1\over \ell} A_0  ~.
\label{chain3}
\eneq
The  velocity $c$ is given by $c = \ell J/\pi\hbar$.   $x_0 = c t$ and
$(A_0, A_1) = (\lambda, c A)$. Although the Maxwell term is absent in the
$\ell \go 0$ limit, it is generated at finite $\ell$.    From the
dimensional analysis $e = k (J/\ell)^{1/2}$ where $k$ is a constant of
O(1).  (\ref{chain3}) is nothing but the two-flavor massless
Schwinger model in an uniform background charge density.

In the  ladder system (\ref{ladder1}) the rung interaction  
gives an additional four-fermi interaction:
\beqn
\L_{\rm ladder} = \sum_{j=1}^{n_\ell} 
   \L_{\rm chain}[\psi_j, A_j] + \L_{\rm int} \hskip 2.8cm &&\cr
\L_{\rm int} =  {\ell J'\over 4} \sum_{(jk)} \int dx \, \Big(
\{ \psi_j^\dagger \psi_k^{} , \psi_k^\dagger \psi_j^{} \} +
\psi_j^\dagger \psi_j^{} \cdot \psi_k^\dagger \psi_k^{} 
     \hskip .3cm && \cr
+ \{ \psibar_j \psi_k , \psibar_k \psi_j \} + 
\psibar_j \psi_j \cdot \psibar_k \psi_k  \Big) &&
\label{ladder2}
\eeqn
where every quantity in the expression is a flavor singlet.

The continuum theories (\ref{chain3}) and (\ref{ladder2}) are solved
by bosonization.  On a circle fermion operators are bosonized
in the interaction picture by\cite{qed1,qed2}
\beqn
\psi^a_\pm(t,x) = {\mybig 1\over\mybig \sqrt{L}} 
\, C^a_\pm \,
 e^{\pm i \{ q^a_\pm(t) \pm 2\pi p^a_\pm  x/L \} }
  N_0[ e^{\pm i\sqrt{4\pi}\phi^a_\pm (t,x) } ] \quad (a=1,2)
\label{bosonization}
\eeqn
where $L$ is the volume of the circle,  $C^a_+=e^{i\pi
\sum_{b=1}^{a-1}(p^b_++p^b_-)}$  and $C^a_-=e^{i\pi\sum_{b=1}^a(p^b_+-p^b_-)}$.  
$N_\mu[\cdot\cdot]$ denotes the normal ordering with respect to a mass
$\mu$. The only physical degree of freedom associated with the gauge fields
is the Wilson line phase, $\Theta_W = (e/\hbar c)\int_0^L dx\, A_1(t,x)$.
The bosonized Hamiltonian is expressed in terms of zero modes $q_\pm ^a$,
$\Theta_W$, oscillatory modes $\phi^a$, and their canonical conjugates.

 In the chain system
the gauge interaction gives rise to a mass for 
$\Phi=(\phi^1 + \phi^2)/\sqrt{2}$ ; $\mu_\Phi^2=\mu^2 \equiv 2e^2\hbar/\pi
c^3$.  The other  combination $\chi=(\phi^1 - \phi^2)/\sqrt{2}$ remains
massless, which correponds to the gapless excitation in the Bethe ansatz.  

When $|J'/J| \ll 1$, $\L_{\rm int}$ in (\ref{ladder2}) is treated
as a perturbation. In the absence of $\L_{\rm int}$ we have $n_\ell$
massive bosons $\Phi_j$'s  and $n_\ell$ massless bosons $\chi_j$'s.
The question is whether or not all $\chi_j$'s acquire masses due to
$\L_{\rm int}$.  

The first two terms in $\L_{\rm int}$ are vector-like.  They change
the spin wave velocity, but do not generate a gap.
The last two terms in $\L_{\rm int}$ break the chiral
invariance, and therefore  can lead to gap generation.

The gap generation is related to generation of chiral condensates.
It occurs in a self-consistent manner.  In the bosonization language
nonvanishing chiral condensates in the infinite volume limit appear
when all relevant boson fields become massive.  The critical identity
is, for a boson field $\phi$, 
$N_0[e^{i\beta\phi}] = B(\mu L)^{\beta^2/4\pi} N_\mu[e^{i\beta\phi}]$.
Here $B(z) \sim e^\gamma z/4\pi$ for $z\gg 1$.  In other words the factor
$L^{-1}$ resulting from the insertion of (\ref{bosonization}) into
$\psibar\psi$ can be cancelled by the $B(\mu L)$ factor in the $L\go\infty$
limit, provided all relevant $\mu$'s are nonvanishing.   
Boson masses are also determined by (\ref{ladder2}) 
expanded in a power series in boson fields.  Coefficients thus obtained
depend on boson masses to be determined, which leads to self-consistency
conditions.\cite{qed2}

Careful examination shows that the zero modes also play a crucial role
in the determination of boson masses.   In a cyclically symmetric
$n_\ell$-leg ladder system, the singlet combination of the bosons $\Phi$'s
($\chi$'s) has a mass $\mu_{\Phi_+}$  ($\mu_{\chi_+}$), whereas other
$n_\ell-1$ combinations are degenerate and have masses $\mu_{\Phi_-}$  
($\mu_{\chi_-}$).  $\mu_{\Phi_+}$ is unaltered by the interaction.  For
$n_\ell =2$ and $|J'|\ll J$ one finds
\beqn
&&\mu_{\Phi_-}^2 = \mu^2 + F \mu_{\Phi_-} (\mu_{\chi_-} + \mu_{\chi_+}) \cr
\noalign{\kern 5pt}
&&\mu_{\chi_\pm}^2 = F \mu_{\Phi_-}\mu_{\chi_\pm}  ~~,~~ 
\hskip .5cm  F ={e^{2\gamma}\over 4\pi} {|J'|\ell\over \hbar c} 
\label{mass1}
\eeqn
from which it follows that the spin gap is given by
\beeq
\Delta_{\rm spin} = \mu_{\chi_\pm} c^2
\sim {e^{2\gamma}   k \over 2^{3/2} \pi} \,|J'| = 0.36 k |J'| ~.
\label{gap1}
\eneq

In two leg samples, say in SrCu$_2$O$_3$, $J\sim J'\sim$1300K and 
$\Delta_{\rm spin} \sim 420$K.\cite{exp}  In numerical simulations 
$\Delta_{\rm spin} =.41J'$ (.50$J'$) for small $J'/J$
($J'=J$).\cite{numerical}  Our result (\ref{gap1}) is consistent with both
experiments and numerical simulations.  Similar analysis in the
bosonization language has been previously given by Schulz, by Shelton et
al.\ and by Kishine and Fukuyama.\cite{Schulz}

For an even $n_\ell \ge 4$ (\ref{mass1}) is modified to
\beqn
&&\mu_{\Phi_-}^2 = \mu^2 + n_\ell F \mu_{\Phi_-} 
[\mu_{\chi_-} +\mu_{\chi_+}^{2/n_\ell} \mu_{\chi_-}^{1-2/n_\ell}] \cr
\noalign{\kern 5pt}
&&\mu_{\chi_+}^2 =  2 F \mu_{\Phi_-} 
 \mu_{\chi_+}^{2/n_\ell} \mu_{\chi_-}^{1-2/n_\ell} \cr
\noalign{\kern 5pt}
&&\mu_{\chi_-}^2 =     F \mu_{\Phi_-} 
[n_\ell \mu_{\chi_-} +
(n_\ell - 2) \mu_{\chi_+}^{2/n_\ell} \mu_{\chi_-}^{1-2/n_\ell}] 
\label{mass2}
\eeqn
[Note that the formula (\ref{mass2}) double-counts the rung (12) when
$n_\ell=2$.]  The singlet field $\chi_+$ has the smallest mass.

For an odd $n_\ell$, dynamics in the zero mode sector do not allow  a
solution with $\mu_{\chi_\pm} \not= 0$.  The singlet field $\chi_+$ remains
massless.

We have shown that in a cyclically symmetric spin ladder 
system the spectrum becomes gapful if the number of legs is even.
The spin system is mapped to a set of two-dimensional QED models,
which is solved by bosonization.  The gap generation is intimately
connected with chiral symmetry breaking.

\vskip .3cm
\noindent {\it Acknowledgment:} ~
This work was supported in part  by the U.S.\ Department of Energy
under contracts DE-FG02-94ER-40823.

\vskip .3cm
\leftline{\bf References}

\renewenvironment{thebibliography}[1]
	{\begin{list}{\arabic{enumi}.}
	{\usecounter{enumi}\setlength{\parsep}{0pt}
	 \setlength{\itemsep}{0pt}
         \settowidth
 {\labelwidth}{#1.}\sloppy}}{\end{list}}

\def\jnl#1#2#3#4{{#1}{\bf #2} (#4) #3}
\def\em{\it}
\def\nc{\em Nuovo Cimento }
\def\jpA{{\em J.\ Phys.} A}
\def\jpC{{\em J.\ Phys.\ Cond.\ Mat.} }
\def\npB{{\em Nucl.\ Phys.} B}
\def\plA{{\em Phys.\ Lett.} A}
\def\plB{{\em Phys.\ Lett.} B}
\def\prl{\em Phys.\ Rev.\ Lett. }
\def\pr{{\em Phys.\ Rev.} }
\def\prB{{\em Phys.\ Rev.} B}
\def\prD{{\em Phys.\ Rev.} D}
\def\ap{{\em Ann.\ Phys.\ (N.Y.)} }
\def\rmp{{\em Rev.\ Mod.\ Phys.} }
\def\zpC{{\em Z.\ Phys.} C}
\def\sci{\em Science}
\def\cmp{\em Comm.\ Math.\ Phys. }
\def\mplA{{\em Mod.\ Phys.\ Lett.} A}
\def\mplB{{\em Mod.\ Phys.\ Lett.} B}
\def\ijmpB{{\em Int.\ J.\ Mod.\ Phys.} B}
\def\IJMPB{{\em Int.\ J.\ Mod.\ Phys.} B}
\def\ijmpA{{\em Int.\ J.\ Mod.\ Phys.} A}
\def\IJMPA{{\em Int.\ J.\ Mod.\ Phys.} A}
\def\ptp{{\em Prog.\ Theoret.\ Phys.} } 
\def\Zphys{{\em Z.\ Phys.} }
\def\jpsJ{{\em J.\ Phys.\ Soc.\ Japan }}
\def\jmp{{\em J.\ Mod.\ Phys.} }
\def\jssc{{\em J.\ Solid State Chem.\ }}

\def\etal{{\em et al,} }

\vfill


\centerline{\boxit{\hbox{\vtop{\hsize=12.5cm
\small  \noindent 
This article will appear in the Proceedings of {\it SOLITONS}, 
a CRM-Fields-CAP Summer Workshop in Theoretical Physics, 
July 20 - July 26, 1997, Kingston, Ontario, Canada. 
}}}}

\end{document}